\documentclass[manuscript]{aastex61}
\usepackage{amsmath}
\usepackage{graphicx}
\usepackage[outdir=./]{epstopdf}
\usepackage{natbib}
\usepackage{float}
\usepackage{cancel}
\received{}
\revised{}
\accepted{}
\shorttitle{Coronal heating and solar wind formation in QS and CHs}
\shortauthors{Tripathi, Nived and Solanki}
\begin{document} 
\title{Coronal heating and solar wind formation in quiet Sun and coronal holes: a unified scenario}
\author[0000-0003-1689-6254]{Durgesh Tripathi}
\affil{Inter-University Centre for Astronomy and Astrophysics, Post Bag - 4, Ganeshkhind, Pune 411007, India}
\author[0000-0001-6866-6608]{V. N. Nived}
\affil{Armagh Observatory and Planetarium, College Hill, Armagh BT61 9DG, UK}
\affiliation{Inter-University Centre for Astronomy and Astrophysics, Post Bag - 4, Ganeshkhind, Pune 411007, India}
\author[0000-0002-3418-8449]{Sami K Solanki}
\affil{Max-Planck Institute for Solar System Research, Justus-von-Liebig-Weg 3, D-37077, G\"ottingen, Germany}
 \affil{School of Space Research, Kyung Hee University, Yongin, Gyeonggi-Do, 446-701, Republic of Korea}
\begin{abstract}
Coronal holes (CHs) are darker than quiet Sun (QS) when observed in coronal channels. This study aims to understand the similarities and differences between CHs and QS in the transition region using the \ion{Si}{4}~1394~{\AA} line recorded by the Interface Region Imaging Spectrograph (IRIS) by considering the distribution of magnetic field measured by the Helioseismic and Magnetic Imager (HMI) onboard the Solar Dynamics Observatory (SDO). We find that \ion{Si}{4} intensities obtained in CHs are lower than those obtained in QS for regions with identical magnetic flux densities. Moreover, the difference in intensities between CHs and QS increases with increasing magnetic flux. For the regions with equal magnetic flux density, QS line profiles are more redshifted than those measured in CHs. Moreover, the blue shifts measured in CHs show increase with increasing magnetic flux density unlike in the QS. The non-thermal velocities in QS, as well as in CHs, show an increase with increasing magnetic flux. However, no significant difference was observed in QS and CHs, albeit a small deviation at small flux densities. Using these results, we propose a unified model for the heating of the corona in the QS and in CHs and the formation of solar wind.
\end{abstract}
\section{Introduction} \label{sec:intro}
The solar surface at coronal temperatures is usually covered with three types of large-scale features, namely, Coronal Holes (CHs), Quiet Sun (QS) and active regions (ARs). CHs appear darker, relative to the surrounding QS regions, at radiation formed at coronal temperatures at EUV and X-ray wavelengths \citep[see\textit{,} e.g.,][]{chdef, StuSSR_2000, StuSP_2002}. CHs have been often studied partly because they are the source regions of fast solar wind \citep[see\textit{,} e.g.,][]{HasDL_1999, TuZM_2005, HeiJT_2020}.

In radiation coming from the photosphere and the chromosphere, on-average there is hardly any noticeable difference in the intensities except in the observations taken in spectral lines of \ion{He}{1}~10830~{\AA} and \ion{He}{1}~{584}~{\AA}. We note that the \ion{He}{1}~10830~{\AA} is an absorption line whereas \ion{He}{1}~{584}~{\AA} is an emission line. In images recorded using \ion{He}{1}~10830~{\AA}, CHs appear brighter than QS \citep[see\textit{,} e.g.,][]{ch1083}, whereas those taken in \ion{He}{1}~{584}~{\AA} appear about 30\% darker \citep[][]{WilLD_1998, JorMS_2001}. However, such differences could be attributed to the sensitivity on the formation of \rm{He} lines on to the EUV radiation as well as energetic electrons from the corona, so that both lines are weakened \citep[see, e.g.,][]{AndJ_1997, CenTUC_2008, LeeGC_2016}. At radio wavelengths (1.21 mm), however, CHs are indistinguishable from QS \citep{BraSB_2018}. It is also found that there is hardly any difference in the distribution of magnetic flux density in solar photosphere in QS and CH \citep[][]{KayTSP_2018}.

It is widely believed that the structure of the magnetic field in CHs is different from that in QS, although there is hardly any difference in the distribution of photospheric magnetic flux density \citep[see, e.g.][]{KayTSP_2018}. In CHs, a large fraction of field lines is open and extend into interplanetary space \citep{chopen} allowing hot plasma to escape. Indeed, spectroscopic observations recorded by the Coronal Diagnostics Spectrometer \citep[CDS;][]{CDS} and Solar Ultraviolet Measurement of Emitted Radiation \cite[SUMER;][]{SUMER} onboard the Solar and Heliospheric Observatory (SOHO) show blue shifts in addition to reduced intensities in coronal lines \cite[][and references therein]{WilLD_1998, StuSRSBSWH_1999, Peter_1999, StuSSR_2000, StuSP_2002, XiaMC_2003}. 

\cite{StuSP_2002} studied the properties of 14 ultraviolet emission lines covering the temperature range of 3.9$\times$10$^{4}$~K to 2.25$\times$10$^{6}$~K. While all these lines were blue-shifted inside the CHs relative to QS, only the spectral lines with peak formation temperatures larger than 7$\times$10$^{5}$~K clearly showed reduced intensities in the CHs, in agreement with the results obtained by \cite{HubFN_1974} using \textsl{Skylab} observations. \cite{StuSP_2002} also showed a positive correlation between blue shifts and line intensities in transition region line in CHs, similar to the result obtained by \cite{HasDL_1999} and  \cite{StuSSR_2000}. Moreover, these authors found that the spectral lines with a temperature lower than 4$\times$10$^{5}$ had larger width in CHs than in QS, whereas the high-temperature lines were narrower.

Such observations lead to the question at what temperature the CHs and QS start becoming distinct and what may be the possible cause. An answer may not only provide new insights into the formation and acceleration of solar wind in coronal holes \citep[see\textit{,} e.g.,][]{HasDL_1999, StuSSR_2000, TuZM_2005} but also the flow of mass and energy within the solar atmosphere.

Combining the above mentioned observations with modelling of the magnetic field and using Rosner-Tucker-Vaiana \citep[RTV;][]{rtv} scaling laws, \cite{qsbright} proposed that the observed difference between the QS and CH at coronal temperatures could be attributed to the density of coronal loops with different lengths and heights. According to their extrapolations, CH and QS have different densities of higher and longer loops but a similar number of low-lying, shorter loops. This could explain the lower intensities in coronal lines in CHs, but similar intensities in the transition region and cooler radiation, although they found somewhat fewer intermediate and shorter loops in CHs, suggesting a difference in brightness between CHs \& QS also in the TR. Such a difference in the TR has so far not been properly established.

It is well known that radiances observed in UV spectral lines are highly dependent on the amount of magnetic flux present. Therefore, it is imperative to study the radiance as well as other line parameters such as Doppler shifts and non-thermal widths by taking into account the distribution of magnetic flux, which may be different in QS and CH. 

In this paper, for the first time, we perform such an analysis for the transition region \ion{Si}{4} line at 1393~{\AA} recorded by the Interface Region Imaging Spectrograph \citep[IRIS,][]{iris}. Here, we take the magnitude of magnetic flux density into account when comparing the line properties in transition regions (following the study by \cite{KayTSP_2018} for the \ion{Mg}{2}~k line). We compare the line's intensity, Doppler shift and non-thermal width at similar magnetic flux density in QS and CH. The remainder of the paper is structured as follows. We present our observations and data in \S\ref{sec:method} followed by analysis and results in \S\ref{sec:res} \& \S\ref{final_results}. We summarise, discuss and conclude in  \S\ref{sec:sum}.

\section{Observations and Data}\label{sec:method}

\begin{table}
\centering
\caption{Details of IRIS rasters used in this study. The exposure time for each study is 31.6~s.}
\begin{tabular}{ |c|c|c|c|c|c|c| }
\hline
Data    & Date of           	& Start  time   	& End time     	& FOV 		& $\mu$-value &Observation ID\\
        	& Observation       & (UT)       		&(UT)			& (arcsec) 	&				& (OBSID)            \\
\hline
Set 1 &2014/07/26& 00:10:28& 03:40:53&$141''\,\times174\,''$ &0.85 &3824263396 \\
Set 2 &2014/07/24& 11:10:28&14:40:53&$141''\,\times174\,''$ &0.97&3824263396  \\
Set 3 &2015/10/14& 11:07:33&14:37:58&$141''\,\times174\,''$ &0.95&3820263396  \\
\hline
\end{tabular}
\label{tab1}
\end{table} 

We use data from the IRIS spacecraft, which obtains spectra in three wavelength bands, namely 1332{--}1358~{\AA}, 1389{--}1407~{\AA} and 2783{--}2834~{\AA} with an effective spectral resolution of 26 m{\AA} and 53 m{\AA}, respectively \citep{iris_tech}. The effective spatial resolution of the far UV spectra is 0.33{\arcsec} while of the near UV spectra is 0.4{\arcsec}. The strongest transition region lines recorded by IRIS are \ion{Si}{4}~1393~{\AA} and 1402~{\AA}. For this study, we have used the 1393~{\AA} line that,  according to theory, is expected to be a factor of two stronger than 1402~{\AA} in optically thin conditions  \citep[see however,][]{GonV_2018, TriNID_2020}. The stronger line was chosen to achieve a better signal-to-noise ratio.

For this study, we have analyzed three sets of observations, the details of the which are presented in Table~\ref{tab1}. In keeping with our goal we have selected observations that contained QS as well as CH in the same raster. This is to avoid any bias due to observations taken at different locations and different times, as that may lead to differences in the results due to different angle to the solar surface as well as instrument degradation. For our work, we have used the level-2 data that are corrected for instrumental effects and orbital variations etc.

For the magnetic field measurements, we have used the line-of-sight (LOS) magnetograms recorded by the Helioseismic and Magnetic Imager \citep[HMI,][]{hmi} onboard the Solar Dynamics Observatory \citep[SDO,][]{sdo}. We have further used the observations at 1600~{\AA} \& 193~{\AA} recorded by the Atmospheric Imaging Assembly (AIA), also on board SDO, at corresponding times. The later channel is used for the identification of coronal holes, while the former is used for co-alignment of the IRIS observations with AIA and HMI observations. 

AIA provides full disk images of the Sun in 8 UV/EUV channels with a pixel size of about 0.6{\arcsec} and temporal resolution of 12~s for EUV and 24~s for UV channels \citep[][]{aia2, aia, aia3}. The HMI provides measurements of the photospheric magnetic field with a cadence of 45~s (LOS) and 720~s (vector) and pixel size of 0.5\arcsec. For details see \cite{hmi, hmi2}. 

\section{Data Analysis and Results}\label{sec:res}

Since the methods of data analysis are exactly the same for all the three cases, below we detail those for one data set in \S\ref{data1}. The plots and results for the other two sets are shown in Appendix~\ref{sec:app_a} \& \ref{sec:app_b}, respectively. Finally, in order to have better statistics, we have obtained an average of all three observations and presented the final results in \S\ref{final_results}. The addition was possible because all three observations were taken at very similar $\mu$ values.

\subsection{The dataset 1: observation on July 26, 2014}\label{data1}

\begin{figure*}
\centering
\includegraphics[width=0.95\textwidth]{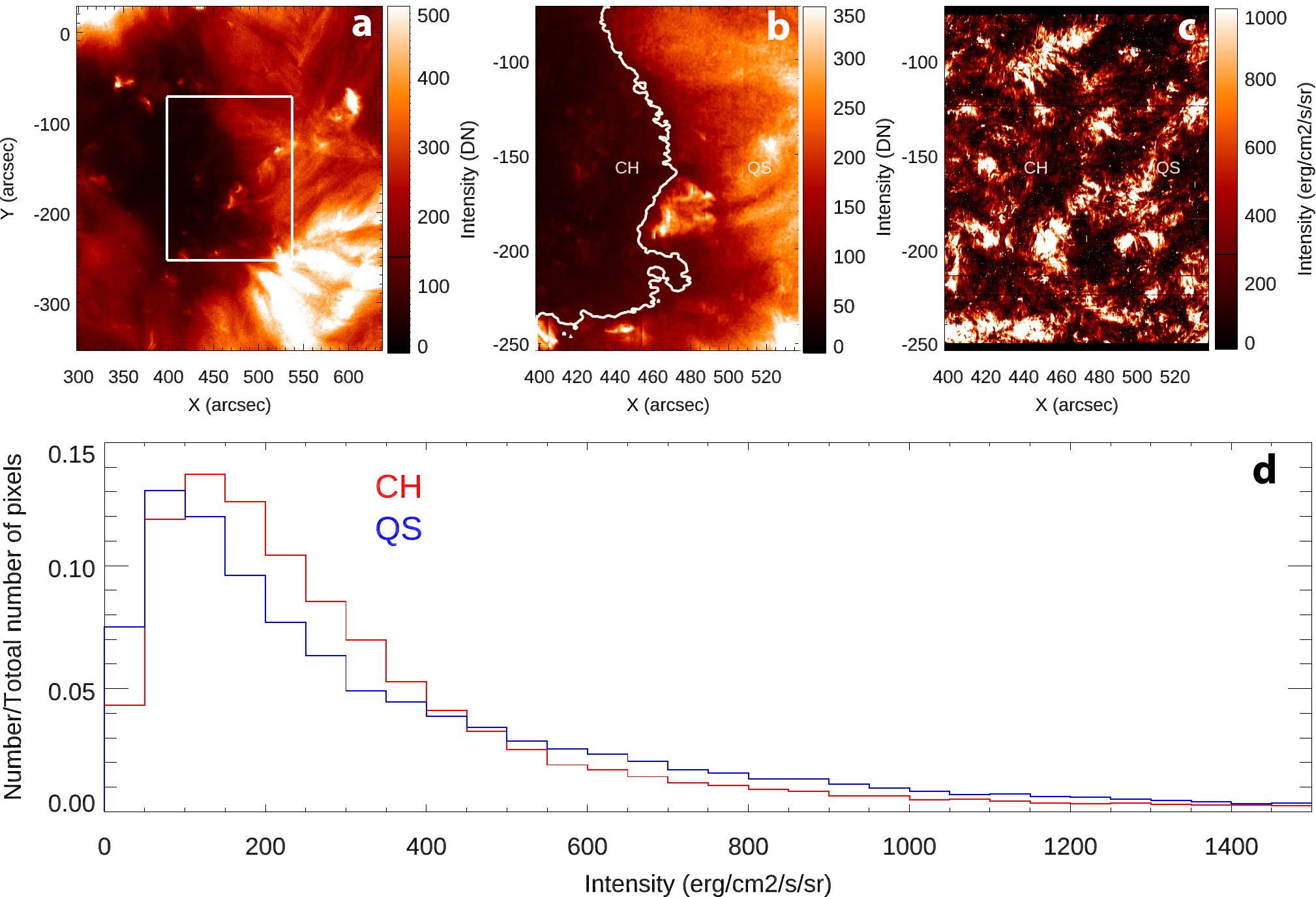}
\caption{Coronal and transition region observation of coronal hole (CH) and quiet Sun (QS). Panel (a): A portion of the Sun's disk observed by AIA using the 193~{\AA} channel. The over-plotted box depicts the IRIS raster field-of-view (FOV). Panel (b): Rastered AIA 193~{\AA} image restricted to the IRIS FOV over-plotted with contours of 80~DN to distinguish between the CH and QS. Panel (c): Intensity map obtained in \ion{Si}{4}~1394~{\AA} using IRIS raster. Panel d: Histograms of the intensities for QS (blue) and CH (red).}
\label{fig:fig1}
\end{figure*}

A portion of the Sun's disk recorded on July 26, 2014, with AIA's 193~{\AA} channel is shown in the panel (a) of Fig.~\ref{fig:fig1}. The over-plotted box shows the FOV of the IRIS raster. We display an AIA~193~{\AA} image corresponding to the IRIS FOV in panel (b). We define CHs (QS) as the region with intensity values lower (higher) than 80~DN (data number) in AIA 193~{\AA} channel (the final boundary was set after manual inspection). A contour of 80~DN is over-plotted on the 193~{\AA} images (see panel b). To identify the boundary between CHs and QS, we have used coronal images recorded by AIA in 193~{\AA}. Since the IRIS raster was taken for a duration of $\sim$3.5 hrs, a single AIA image can not be used. Therefore, we created a synthetic rastered AIA~193~{\AA} image covering the IRIS raster duration by combining parallel strips in AIA~193~{\AA} images taken every 10 mins. We then identify the boundary of CH \& QS in this rastered image. Note that to keep the effect of solar rotation, which is present in the IRIS rasters, the rastered 193~{\AA} image was obtained without compensating for the solar rotation. We display the intensity map obtained in \ion{Si}{4}~1393~{\AA} in panel (c). The AIA and IRIS observations together demonstrate that while there is a clear difference in the intensities between CHs and QS at coronal temperature, the \ion{Si}{4} line, mapping the transition region, does not reflect any conspicuous differences. In panel (d), we plot the distribution of \ion{Si}{4} intensities in QS (blue) and CHs (red), which shows similar characteristics, although with on average somewhat larger brightness in the QS.

\begin{figure*}
\centering
\includegraphics[width=0.95\textwidth]{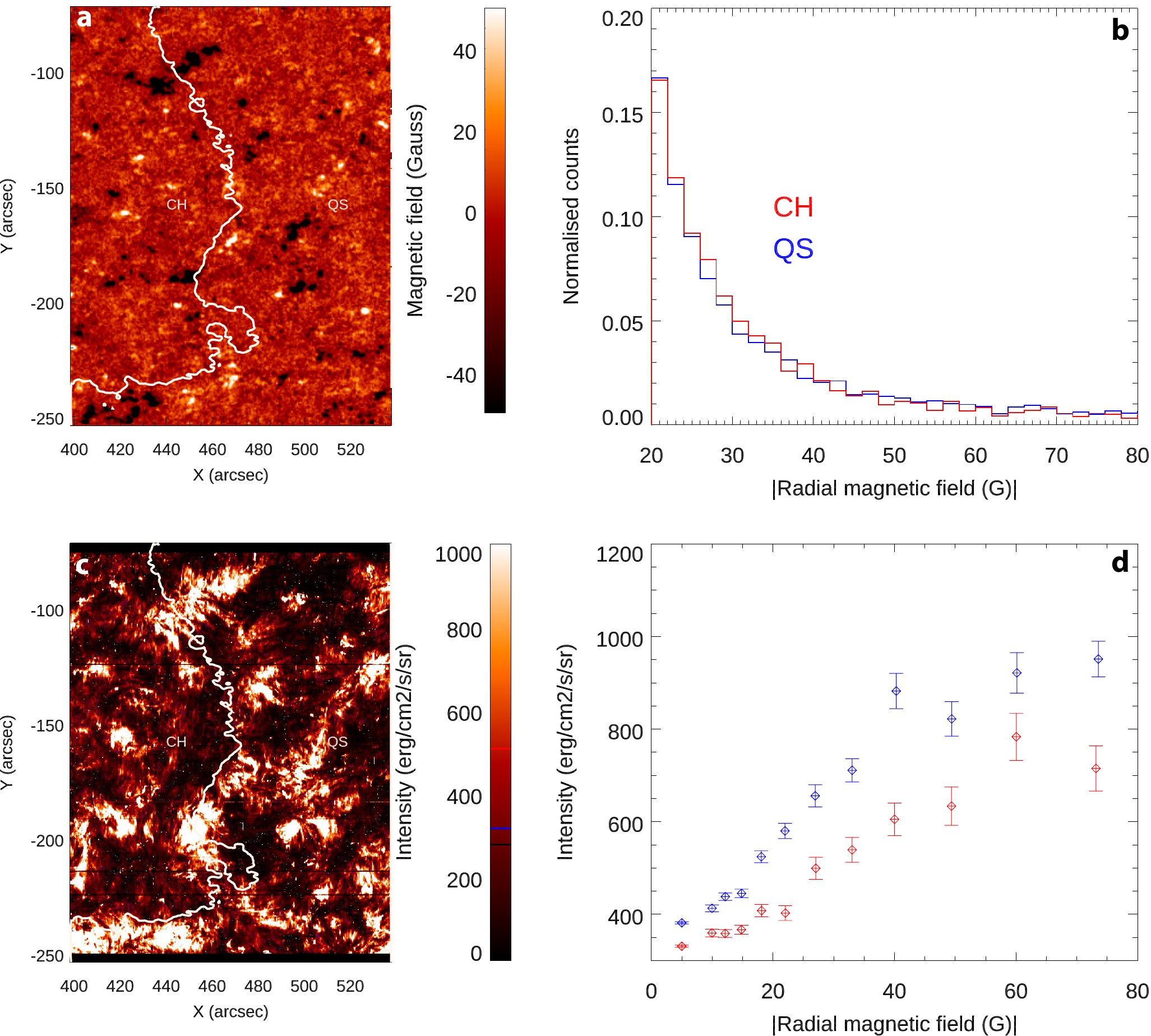}
\caption{Panel a: HMI LOS rastered magnetogram. Over-plotted contour corresponds to 80~DN obtained from AIA~193~{\AA}. Panel b: Histogram of the absolute radial absolute magnetic flux density larger than 20~G in QS (blue) and CH (red). Panel c: Intensity image of the IRIS raster over plotted with AIA~193~{\AA} intensity contours corresponding to 80~DN. Panel d: Average \ion{Si}{4}~1394~{\AA} intensity in CH (red) and QS (blue) as a function of absolute radial magnetic flux density. The error bars represent the standard errors.} \label{fig:fig2}
\end{figure*}

To avoid a bias introduced by different magnetic flux distributions in QS and CH, we compare spectral line properties of \ion{Si}{4} in pixels of identical magnetic flux density in CH and QS. Therefore, it is imperative to co-register and co-align the observations taken from IRIS and HMI as accurately as possible. For this purpose, first we have co-aligned IRIS slit-jaw images obtained at 1400~{\AA} with 1600~{\AA} images recorded by AIA as they both show similar structures. The magnetograms taken with HMI were then co-aligned with 1600~{\AA} images. Co-alignment was performed using the standard SSW routines \textsl{coreg\_map.pro} and \textsl{correl\_offset.pro}, which use the cross-correlation to align the images. Using the time series of co-aligned HMI magnetograms, we obtained a rastered magnetogram that corresponds to the IRIS raster. Similar to AIA~{193}~{\AA} rastered image, the rastered magnetogram was also obtained without compensating for the solar rotation. Note that for this purpose we have used the HMI images taken every three minutes. The rastered magnetic field map is shown in Fig.~\ref{fig:fig2}a. The over-plotted contours represent the 80~DN obtained from AIA~193~{\AA}.  We also plot the distribution of the unsigned magnetic flux density for QS (blue) and CH (red) in Fig.~\ref{fig:fig2}a. To avoid the errors, we have only plotted the distribution for unsigned magnetic flux density larger than 20~G \citep[][]{YeoFSC_2014}. We note the almost identical distribution of flux density in CH and QS, similar to \cite{KayTSP_2018}.

Using the artificially rastered LOS magnetogram, we obtain the radial component of the magnetic field at each pixel (which is expected to be the dominant component for kG fields; \cite{solanki_1993}) by dividing the LOS component by the corresponding $\mu$-values. Finally, we study the variation of the intensities of the \ion{Si}{4} line as a function of absolute magnetic flux density. 

For this purpose, we first created a scatter plot and then grouped pixels based on their absolute radial magnetic flux density within bins with a constant logarithmic width of 0.2. The binning improves the statistics at higher magnetic flux densities. Since the estimated error on the HMI LOS field measurements is $\pm$10~G \citep[see, e.g.][]{YeoFSC_2014}, we have assigned a value of 5~G to each pixel with absolute magnetic flux density less than 10G.

We averaged the intensity and absolute radial magnetic flux density in each magnetic field bin and plotted the intensity as a function of the radial magnetic field in Fig.~\ref{fig:fig2}d. In the plot, blue is for QS and red is for CH. The error bars represent the standard error \footnote{Standard error is defined as the ratio between the standard deviation and square root of the sample size.}. The larger error bars at higher magnetic field reflect poorer statistics. For example, we have 104 (232) and 86 (220) data points in the last two bins ($|B|>55~G$) for CH (QS). The plot demonstrates that the average QS intensity in each radial magnetic field bin is larger than that in CH and that the difference in the intensities tends to increase with the increasing absolute magnetic flux density. 

The \ion{Si}{4} observations \citep{FraR_1977} recorded by OSO-8 satellite did not show any difference in the line intensities integrated over QS and CHs. However, our results are in agreement with those obtained by \cite{StuSP_2002}, which shows that the intensities in spectral lines of \ion{O}{3} ($\log\,T=4.9$), \ion{O}{4} ($\log\,T=5.15$), and \ion{O}{5} ($\log\,T=5.35$) are reduced in the coronal holes relative to QS.

As stated earlier, we have studied three data sets. We note that the other two data sets give similar results and figures are shown in  Appendices. 

\subsection{Doppler Velocities: Upflows and downflows}
\begin{figure*}
\centering
\includegraphics[width=0.95\textwidth]{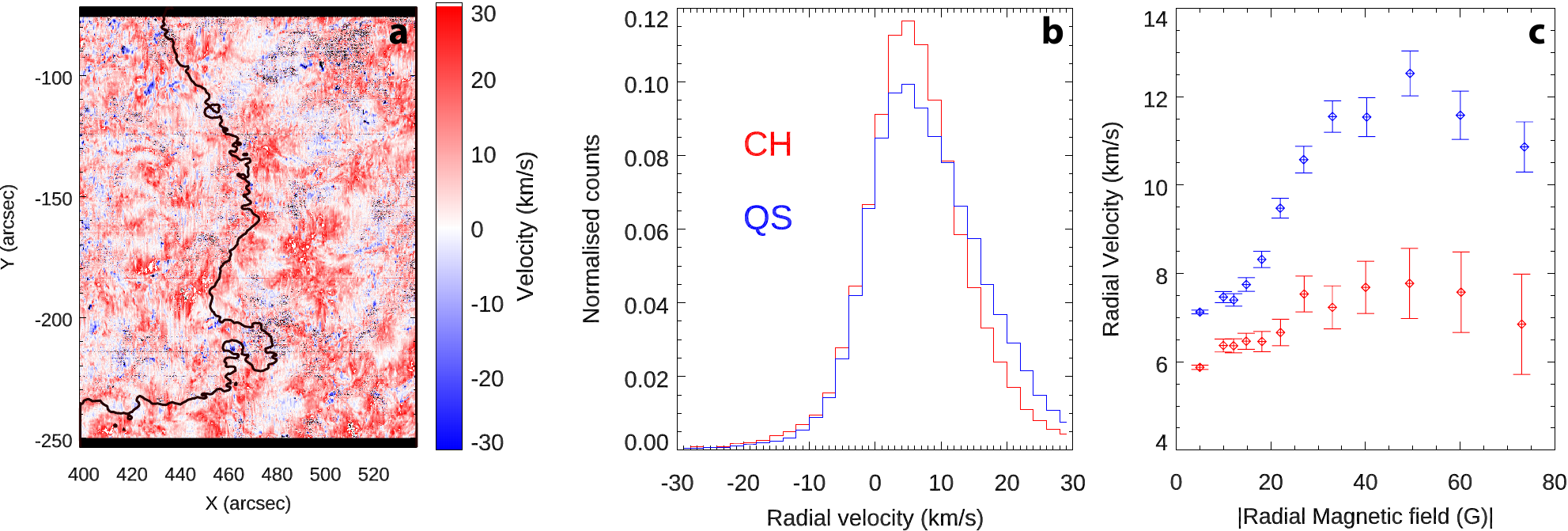}
\caption{Doppler velocity in CHs and QS. Panel a) Doppler velocity map obtained in  \ion{Si}{4}~1394~{\AA} over-plotted with contours of AIA~193{\AA} intensities corresponding to 80~DN, which mark the boundary of the CH. Panel b) Histograms of velocities in CHs (red) and QS (blue). Panel c) Velocity in CHs (red) and QS (blue), averaged over bins, as a function of absolute radial magnetic flux density.}\label{fig:fig3}
\end{figure*}

For the wavelength calibration needed to study the Doppler shift in CHs and QS using the \ion{Si}{4}~1394~{\AA} line we have used \ion{O}{1}~1355.60~{\AA} with peak formation temperature at $\log\, T [K]= $4.5 Since the spectral lines corresponding to transitions of neutral atoms or of singly ionized atoms are expected to be close to rest \citep{HasRO_1991}, the \ion{O}{1} line, therefore, can be used to derive an absolute wavelength calibration. By fitting a single Gaussian to the observed \ion{O}{1} spectra and thereby obtaining the centroid, we obtain the reference wavelength for \ion{Si}{4}~1394 {\AA} denoted by $\lambda_{ref\_SiIV}$ as,

\begin{align}
\lambda_{ref\_SiIV} &= \lambda_{SiIV\_lab} - (\lambda_{OI\_lab}-\lambda_{OI\_obs}),
\end{align}

\noindent where $\lambda_{SiIV\_lab}$  is the laboratory wavelength of \ion{Si}{4}~1394~{\AA} line (1393.755 {\AA}) and $\lambda_{OI\_lab}$ is that of the \ion{O}{1} line (1355.598 {\AA}), which are taken from \citep{sandlin}. We display the Doppler velocity map obtained for \ion{Si}{4}~1394~{\AA} in Fig.~\ref{fig:fig3}.a. The over-plotted contours correspond to 80~DN on AIA~193~{\AA}. Similar to the intensity measurements, there are no clear visual differences in the Doppler shift patterns of QS and CHs. Panel (b) shows that the histogram of LOS velocities is higher for CHs (blue) than for QS (red) within the range of $-$30 to about 10~km~$s^{-1}$, while at larger redshifts, the QS histogram dominates. We note that, there are only 19\% (17\%) blue shifted pixels in CHs (QS).

We now compare the Doppler velocities in CHs and QS for pixels with near-identical magnetic flux density, following the procedure used for intensities, except that the signed averaged velocities in each bin are plotted versus absolute magnetic flux density. Note that we have divided the measured Doppler velocities at each pixel with the corresponding $\mu$-values to obtain the radial speed of the flows. The plot clearly reveals that the average velocities in CHs and QS are positive independent of the magnetic flux (see Fig~\ref{fig:fig3}.c). However, on average, the line is more red-shifted in the QS than in CHs. We further note that the difference in averaged velocities increases with increasing absolute magnetic flux density. 

\subsection{Non-thermal Velocity}
The observed width of a spectral line is produced by the combined effect of thermal, non-thermal and instrumental broadening (neglecting the generally small natural width). The full-width-at-half-Maximum (FWHM, $\delta\lambda$) of the observed line is given by 

\begin{align}
\delta\lambda=\frac{\lambda_{0}}{c}\sqrt{4~ln2\left(\frac{2K_BT_i}{m}+\epsilon^2\right)+\sigma_I^2}, \label{eqn:eqn1}
\end{align}

\noindent where $\lambda_0$ is the laboratory wavelength, $K_B$ is the Boltzmann constant, $T_i$ is the ion temperature (63000~K), $m$ is the mass of the ion (28.085 amu (g/mol)), $\epsilon$ is non-thermal velocity, and $\sigma_I$ is the instrument spectral broadening. IRIS $\sigma_I$ for FUV spectra is 31.8 mA \citep{inst_width}. 

\begin{figure*}
\centering
\includegraphics[width=0.95\textwidth]{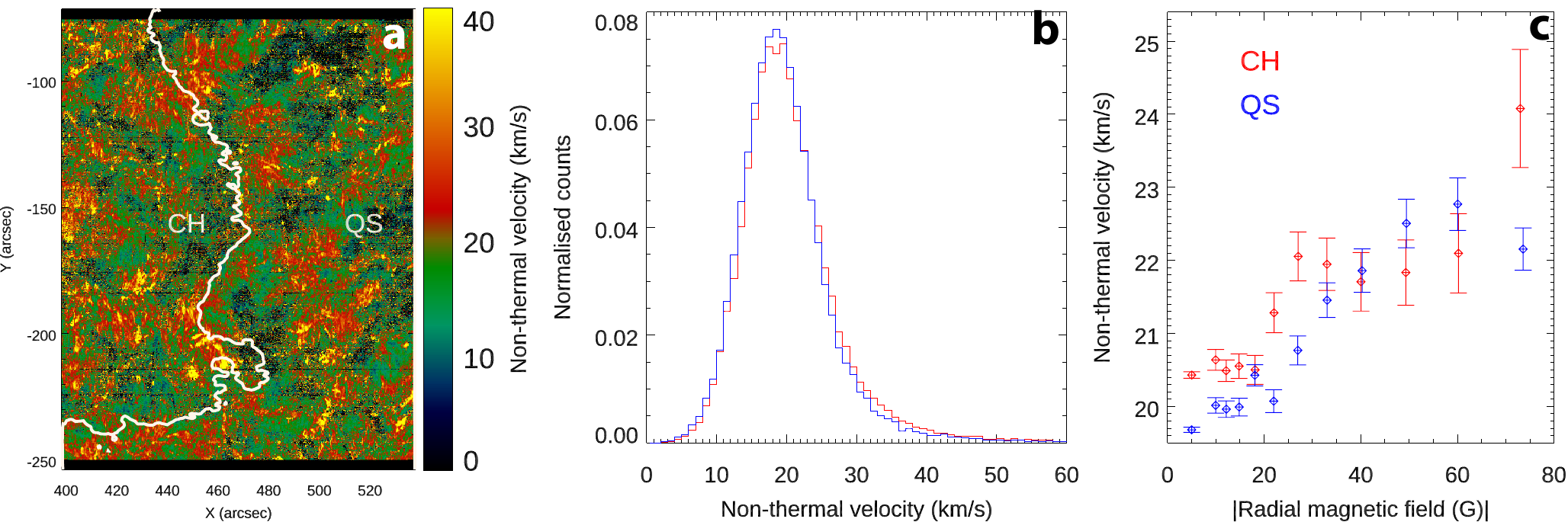}
\caption{Panel a: Non-thermal velocity map obtained in the \ion{Si}{4}~1394~{\AA} line over-plotted with contours marking the boundary of the CH. Panel b: Histograms of non-thermal velocities in CHs (red) and QS (blue). Panel c: Average non-thermal velocity in CH (red) and QS (blue) as a function of absolute radial magnetic flux density.}\label{fig:fig4}
\end{figure*}

We display the obtained maps of $\epsilon$ in panel a of Fig.~\ref{fig:fig4}a. The over-plotted contours are the 80~DN of AIA 193~{\AA}. We show the histograms of $\epsilon$ in CH (red) and QS (blue) in panel b, which suggest a slightly larger value in CH with respect to QS. As was the case for intensity and Doppler maps, the non-thermal maps also do not show any obvious visual difference between QS and CHs. We plot the averaged non-thermal width for QS (blue) and CH (red) as a function of the absolute magnetic flux density in panel c. Note that we have used the same magnetic field bins as for intensity and Doppler shift. For both CH and QS $\epsilon$ clearly increases with increasing radial magnetic field strength. At lower magnetic flux densities, the \ion{Si}{4} line has slightly larger ($<$1~km~s$^{-1}$) $\epsilon$ than in QS.

\section{Combination of all the studied datasets}\label{final_results}

We now combine the observations of the three solar regions to achieve improved statistics when studying the variation of intensity, Doppler velocity and non-thermal velocity as a function of absolute magnetic flux density in QS and CHs. Since all the three observations were recorded at similar locations on the disk, averaging results from the three data sets should not distort the results or introduce any artefacts. 

\begin{figure*}
\centering
\includegraphics[width=0.85\textwidth]{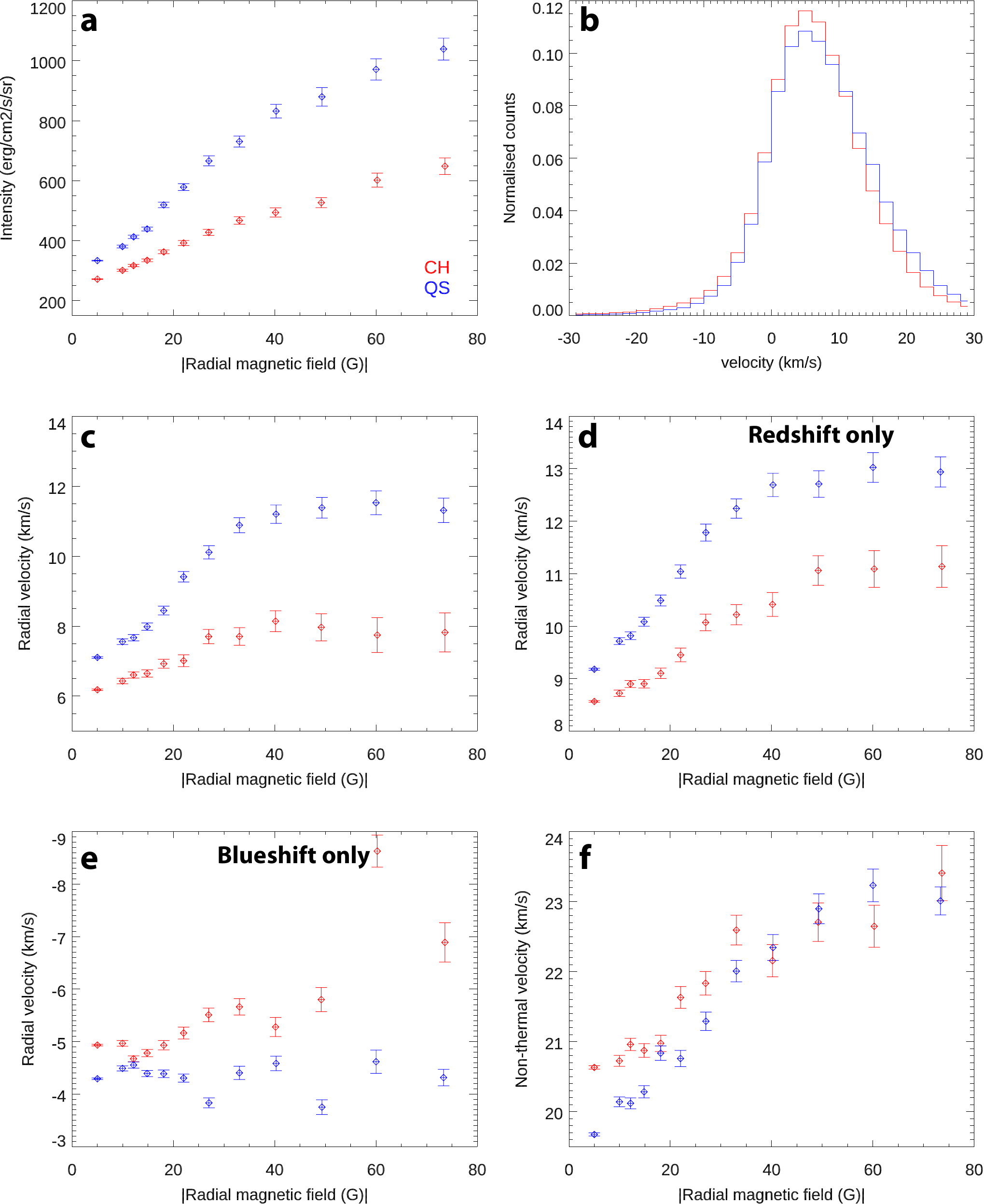}
\caption{Variation of average intensity (panel a), average velocity (panel c) average redshifts (panel d), average blueshifts (panel e) and average non-thermal velocity (panel f) as a function of absolute radial magnetic flux density. Histograms of the velocities are shown in panel (b). }
\label{fig:fig5}
\end{figure*}

In Fig.~\ref{fig:fig5} we display the average intensities (panel a), signed average velocity (panel c) as well as red-shifts (panel d) and blue-shifts (panel e) separately and non-thermal velocity (panel e) versus absolute radial magnetic flux density. As previously done, we have taken a constant bin of 0.2~G of the log of the absolute magnetic flux density. All these plots show smaller error bars due to improved statistics obtained by combining all the three sets of observations. In panel b, we also show the histograms of velocities in the CHs and QS.

Figure \ref{fig:fig5}.a conspicuously displays that the intensity of the QS region is larger than that in CH by a factor of roughly 1.6 when comparing pixels for the regions with the same magnetic flux density. Moreover, the difference in intensities of QS and CH increases with increasing absolute magnetic flux density, with ratios changing from 1.2 at 5G via 1.4 at 20 G,  to 1.7 at 40 G and saturates thereafter. Fig.~\ref{fig:fig5}. b \& c show that the transition regions of both CHs and QS are red-shifted on average, but with the CH plasma displaying a smaller redshift. The difference is small according to the histogram in Fig. 5b. However, because line shifts at small B values dominate the histogram, it masks the bigger differences seen in stronger magnetic features, which is better seen in Fig. 5c.  Further insight can be obtained from Figs.~\ref{fig:fig5}.d and ~\ref{fig:fig5}.e, which consider red- and blueshifts separately. Fig.~\ref{fig:fig5}.d reveals that, redshifts in red-shifted pixels increase with increasing magnetic absolute flux density in both QS and CHs. A striking difference in behaviour is seen, however, when considering the blue-shifted pixels (Fig.~\ref{fig:fig5}.e). In the CHs, the blueshifts increase with increasing magnetic absolute flux density, but no such trend is visible in the QS \citep[see also][]{BryKM_1996}.

We display the combined plots for non-thermal width in Fig.~\ref{fig:fig5}.f. The most striking feature of this plot is the steady increase of non-thermal width with absolute flux density. The plot further confirms that, at smaller flux densities (up to about 30 G), CHs have slightly larger ($<$1~km~s$^{-1}$) non-thermal velocities than in QS. This is suggestive of larger unresolved flows in the inter-network regions of CHs than in QS.

\section{Summary, discussion and conclusions}\label{sec:sum}

In this paper, for the first time, we have studied the intensity, and Doppler and non-thermal velocity in QS and CH in the \ion{Si}{4} line observed by IRIS as a function of the absolute magnetic flux density. We have used LOS magnetic field measurements provided by HMI onboard SDO. To identify QS and CH, we have used AIA~193{\AA} images and for co-aligning IRIS observations with those from HMI and AIA, we have used 1600~{\AA} recorded by AIA. 

Our results show that although there are no visual differences in QS and CH observed in \ion{Si}{4} line (Fig.~\ref{fig:fig1}.c), the histograms of the intensities (Fig.~\ref{fig:fig1}.d) suggest that on average QS is brighter than CH. This becomes clearer when comparing the intensity at similar magnetic flux densities (see Figs.~\ref{fig:fig2}.d \& \ref{fig:fig5}.a). Similar results were obtained for the chromosphere using the \ion{Mg}{2}~k line observed by IRIS \citep{KayTSP_2018}.  

Doppler measurements do not show any marked difference in the Doppler maps obtained for QS and CH when seen by the eye. On average the \ion{Si}{4} is redshifted in both QS and CH. However, when compared quantitatively, the QS show larger redshifts (Fig.~\ref{fig:fig5}.b~\&~c). The average velocity increases with increasing absolute flux density, which is also true when considering only the pixels displaying redshifts (over 80\% of all pixels) (Fig.~\ref{fig:fig5}.d). However, in blueshifted pixels in CHs the outflowing gas flows even faster with increasing absolute flux density, while in the QS, it tends to flow slower (Fig.~\ref{fig:fig5}.e).

\cite{FraR_1977} studied the same \ion{Si}{4} line using observations recorded by the OSO-8 satellite and found no evidence of differences in integrated line intensities or Doppler shifts in CHs relative to QS, in disagreement with results presented here. However, our results, based on far better statistics and much more sensitive data, are in agreement with those obtained by \cite{StuSP_2002}, which shows that on average CH are dimmer and blueshifted relative to QS in transition region lines such as \ion{O}{3}, \ion{O}{4} and \ion{O}{5}. Of which, \ion{O}{3} can be directly compared with the results presented here as its peak formation temperature is close to that of \ion{Si}{4}. 

The non-thermal velocity obtained for both QS and CH shows an increase with increasing absolute magnetic flux density. We further note that the non-thermal velocities in CHs are slightly larger (by $<$1km~s$^{-1}$) (see Fig.~\ref{fig:fig5}d), although only at pixels with smaller flux densities. Using SUMER data \cite{lemaire_1999} and \cite{indist_ch_qs} showed that the transition region lines formed between $\log\,T=4.8$ and 5.4 have larger non-thermal velocity in CHs relative to QS by 4{--}5~km~s$^{-1}$. Moreover, using data gathered by CDS, \cite{StuSP_2002}, found that transition region lines such as \ion{O}{3}, \ion{O}{4}, \ion{O}{5}, \ion{Ne}{4}, \ion{Ne}{5} and \ion{Ne}{6} have excess width in CHs by 1{--}2~km~s$^{-1}$. These results are in qualitative agreement with our results. Any quantitative discrepancy in the results may be attributed to the different spectral lines observed and the difference in spectral resolution of the three instruments.

Using Rosner-Tucker-Vaiana \citep[RTV]{rtv} scaling laws and extrapolation of photospheric magnetic fields, \cite{qsbright} proposed that the observed differences in the intensities of QS and CH at coronal temperatures could be attributed to the statistics of coronal loops with different lengths and heights. Under this scenario, it was found that QS regions have much higher densities of longer loops than CHs, a somewhat higher density of intermediate length loops that may not reach beyond TR temperatures, but have a similar number of low-lying short loops. Moreover, a similar distribution of photospheric magnetic flux densities in CHs and QS suggests that they consist of a very similar number density of magnetic footpoints. The results obtained in this work show that (1) there is a clear, but not very large difference between CHs and QS at TR temperatures, which may have to do with the small deficit of intermediate length loops in CHs; (2) the main difference between CH and QS appears in regions of comparatively large flux density (except for non-thermal velocities), i.e. in the network. This finding is in line with the fact that the smallest scale loops arise from the inter-network and these are almost equally common in CHs and QS, whereas at least one of the footpoints of bigger loops, which are much less common in CHs than in the QS are located in regions of larger flux, i.e. the network \citep[][]{WieSJ_2010}. These results provide observational support and an extension to the work of \cite{qsbright}.

Taking into account the observational results from this study in combination with those by \cite{HasDL_1999, StuSSR_2000, StuSP_2002} and the magnetic field modelling of \cite{qsbright}, we propose a simple scenario that may qualitatively cover both, the heating of the corona in QS and CHs as well as the formation of the solar wind based on impulsive heating models \citep[see, e.g.][]{Kli_2006}.

Under the ambit of impulsive heating in CHs and QS, there may be a couple of scenarios. Impulsive heating may occur due to magnetic reconnection between closed-closed field lines and closed-open field lines. In the former situation, all that will be observed in \ion{Si}{4} lines is a redshift. Since the total number of closed field lines (both short closed loops and long-closed loops) are smaller in CHs than QS, on average the \ion{Si}{4} line in the QS is expected to be more redshifted, primarily because reconnection between closed-closed field lines will result in other closed field lines. Whereas in case of reconnection between the open-closed field line, some of the plasma may get ejected along the resultant open field and would show a reduced redshift on average. If an impulsive event occurs between closed and open field lines, it can produce jet-like topologies \citep[see, e.g.][]{MorGU_2008}. Jets have often been observed with a cool and a hot component \citep[see, e.g.][]{CanRR_1996, MulDM_2017}. The hot components are interpreted as an evaporation jet, which is produced as a consequence of magnetic reconnection between closed and open field lines \citep[see, e.g.][]{MiyY_2003, MiyY_2004, MorGU_2008}. The cool jet, however, is composed of plasma that is ejected like a slingshot from the kinked location between the closed and open field line \cite[see, e.g.][]{CanRR_1996}. Under this representation, cool jets may be composed of a range of temperatures, depending on the location of the slingshot, so that some cool jets would be observed in \ion{Si}{4}.

The above-presented scenario explains the observational results reasonably well and hints towards a unified picture of heating of the QS corona and of the formation of the solar wind. Moreover, it further suggests that the formation and acceleration solar wind may have started at the height of \ion{Si}{4} line. However, the observations presented here demonstrate that even for the strongest fields, no net blueshift is seen in \ion{Si}{4} averaged over the CHs, although the difference in the redshift of QS and CH is the largest there. In CHs, the redshift at the highest flux densities is larger than that at the smallest flux density. These results suggest that we do not see a clear signature of the formation of the solar wind. The bulk of the solar wind, therefore, must be forming between the heights at which the \ion{Si}{4} and \ion{Ne}{8} lines are formed \citep{TuZM_2005}. According to \citet{TuZM_2005} the acceleration starts somewhere between formation height \ion{C}{4} (5 Mm) to that of \ion{Ne}{8} (20 Mm). They also find upflow speed of 10~km~s$^{-1}$ in \ion{Ne}{8}, whereas we find 6 to 7~km~s$^{-1}$ in \ion{Si}{4}. Under the scenario presented here, this could be explained if the slingshots occur at different heights and accelerate plasma of different temperatures. Only a few occur at \ion{Si}{4} formation heights, most of them higher up. 

We emphasise that this scenario based on magnetic reconnection, which is similar to that suggested by \cite{TuZM_2005}, which in turn is essentially an adaptation of the furnace model of \citet[][]{AxfM_1992}, is not capable of explaining the observed speed of solar wind at 1~AU, lying in the range of 300{--}1000~km~s$^{-1}$. Therefore, further mechanisms are required for the acceleration of the plasma, likely acting at greater distances from the solar surface. The current scenario, however, will facilitate putting the plasma from closed fields to within the open field system.

The scenario presented here may, in principle, also be used to explain the recent discovery of magnetic switchbacks \citep{BalBB_2019, DudKB_2020} in the observations recorded by the FIELDS instrument \citep[][]{BalGH_2016} onboard the Parker Solar Probe (PSP). The resultant field line due to the reconnection between a closed and open field will be ejected with a kink, that may be observed as switchbacks. However, we note that the scenario proposed here is close to the Sun's surface, whereas those observed by FIELDS were taken located 36{--}54~R$_{\sun}$ and it is open whether such switchbacks will survive such a distance. Therefore, further work is required to fully comprehend if such a mechanism is able to explain the new findings by FIELDS.

Although the interpretation provided seems plausible, it requires further analysis of multi-wavelength observations combined with numerical modelling to go beyond the stage of being just a scenario. The spectroscopic observations recorded with the Spectral Imaging of the Coronal Environment \cite[SPICE;][]{spice}, jets at different temperatures followed in Extreme Ultraviolet Imager (EUI; Berghmanns et al. 2019) time series and the measurement of the magnetic field provided by Polarimetric and Helioseismic Imager \citep[PHI;][]{phi} onboard the Solar Orbiter mission \citep{so} in the near future may help us to reach a better understanding.

\begin{acknowledgements} 
We thank the anonymous referee for the careful read of the manuscript and providing useful comments. DT and NVN acknowledge the support by the Max-Planck Partner Group of the Max-Planck Institute for Solar System Research, G\"ottingen. This work has been partially supported by the BK21 plus program through the National Research Foundation (NRF) funded by the Ministry of Education of Korea. DT thanks Nishant Singh for various discussions and comments. IRIS is a NASA small explorer mission developed and operated by LMSAL with mission operations executed at the NASA Ames Research center and major contributions to downlink communications funded by ESA and the Norwegian Space Centre. We would like to thank the AIA, HMI, and IRIS teams for providing valuable data.
\end{acknowledgements}
\bibliographystyle{apj}
\bibliography{ref}
\appendix

\section{The dataset 2: observation on July 24, 2014}\label{sec:app_a}
\setcounter{figure}{0} \renewcommand{\thefigure}{A.\arabic{figure}} 
\begin{figure*}[h]
\centering
\includegraphics[width=0.95\textwidth]{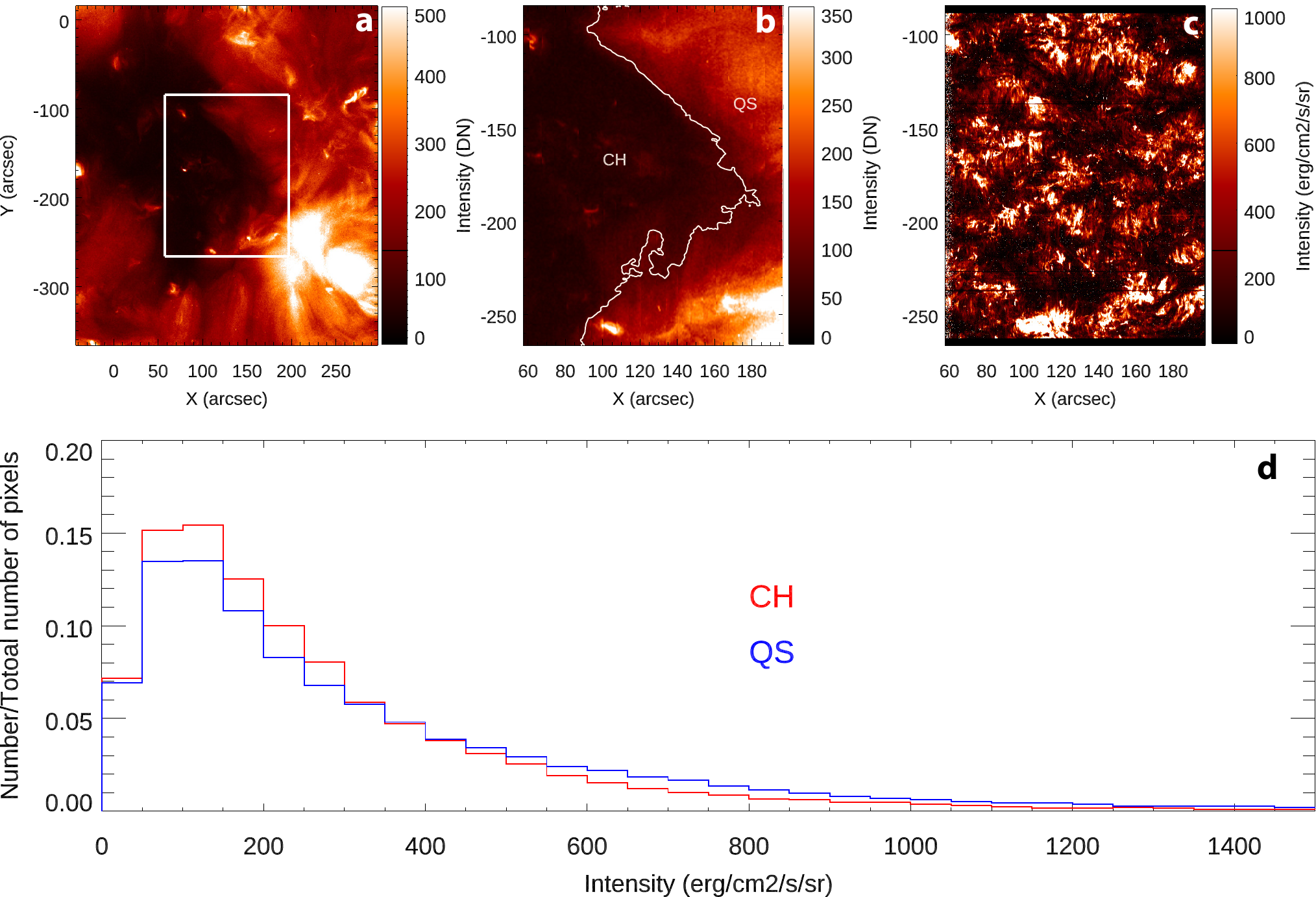}
\caption{Same as Fig.~\ref{fig:fig1} but for observations taken on 24-July-2014.}
\label{fig:fig1_!}
\end{figure*} 
\begin{figure*}[h]
\centering
\includegraphics[width=0.95\textwidth]{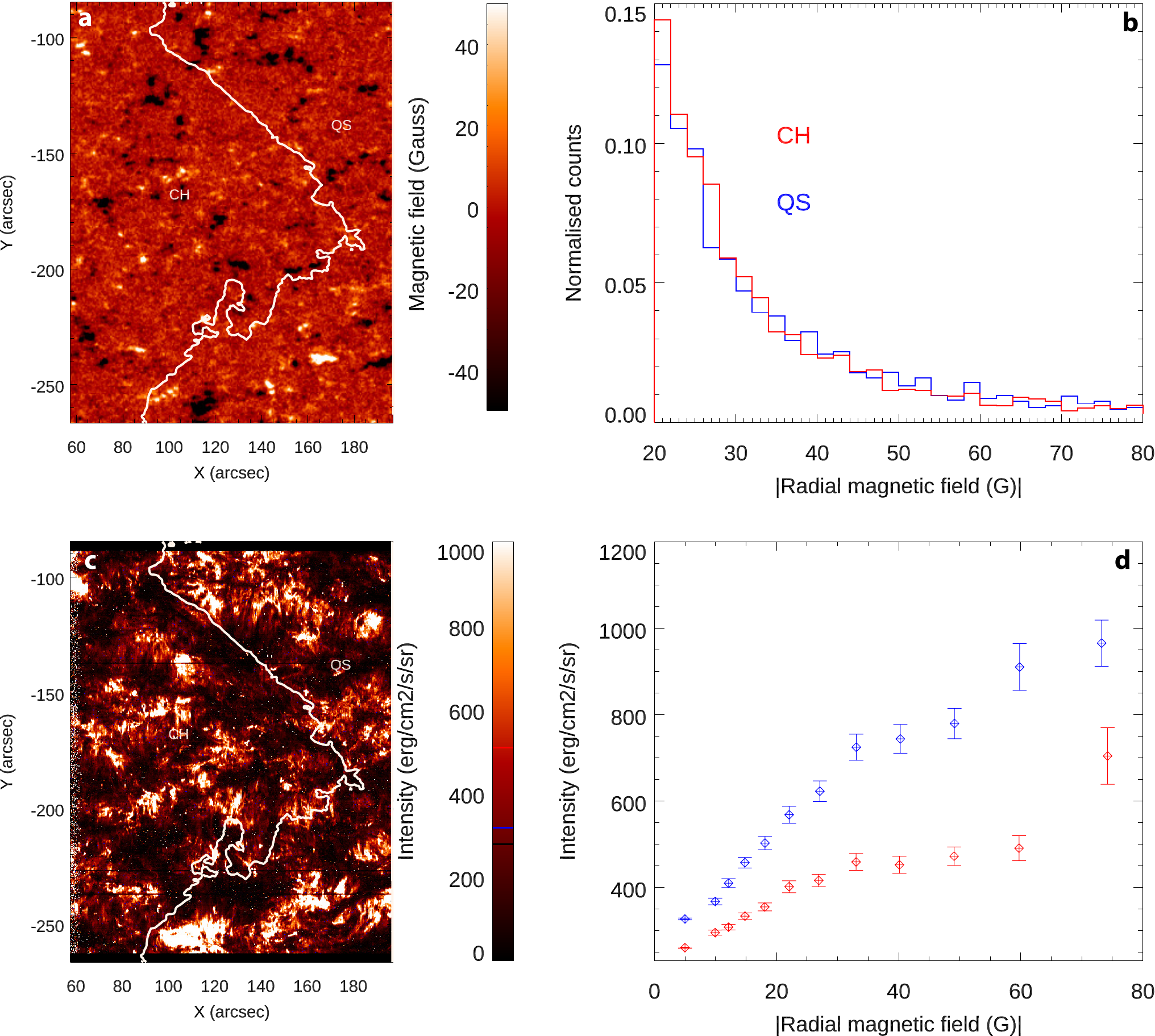}
\caption{Same as Fig.~\ref{fig:fig2} but for observations taken on 24-July-2014.} \label{fig:fig2_1}
\end{figure*}
\begin{figure*}[h]
\centering
\includegraphics[width=0.95\textwidth]{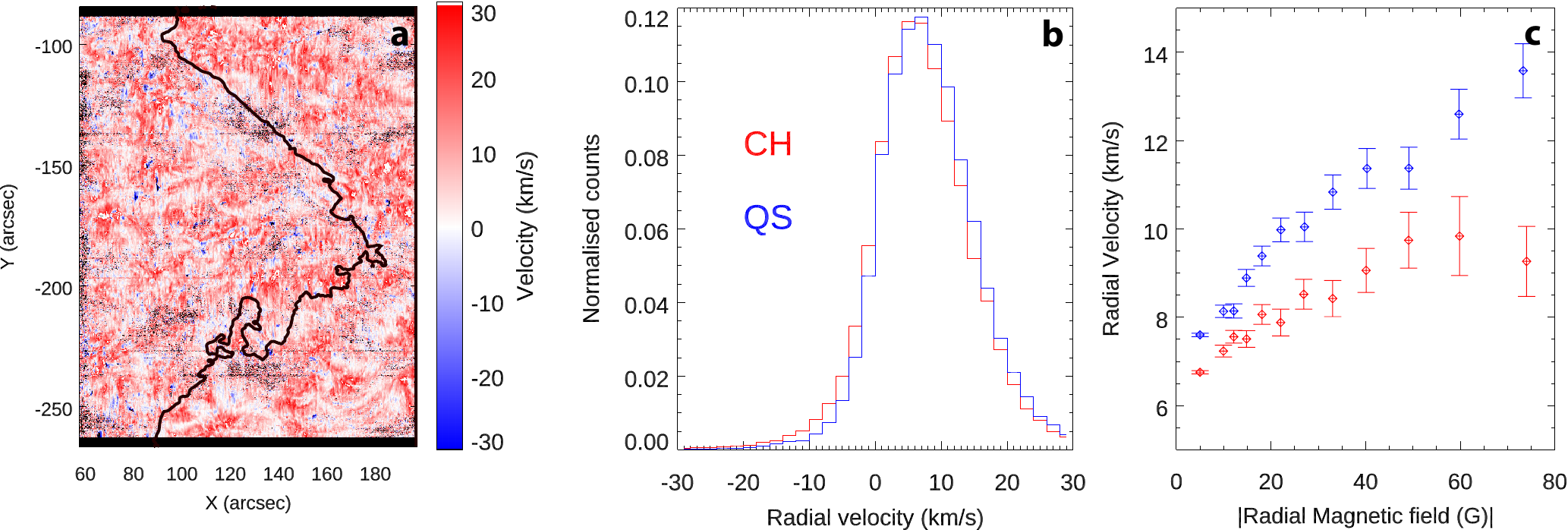}
\caption{Same as Fig.~\ref{fig:fig3} but for observations taken on 24-July-2014.}\label{fig:fig3_1}
\end{figure*}
\begin{figure*}[h]
\centering
\includegraphics[width=0.95\textwidth]{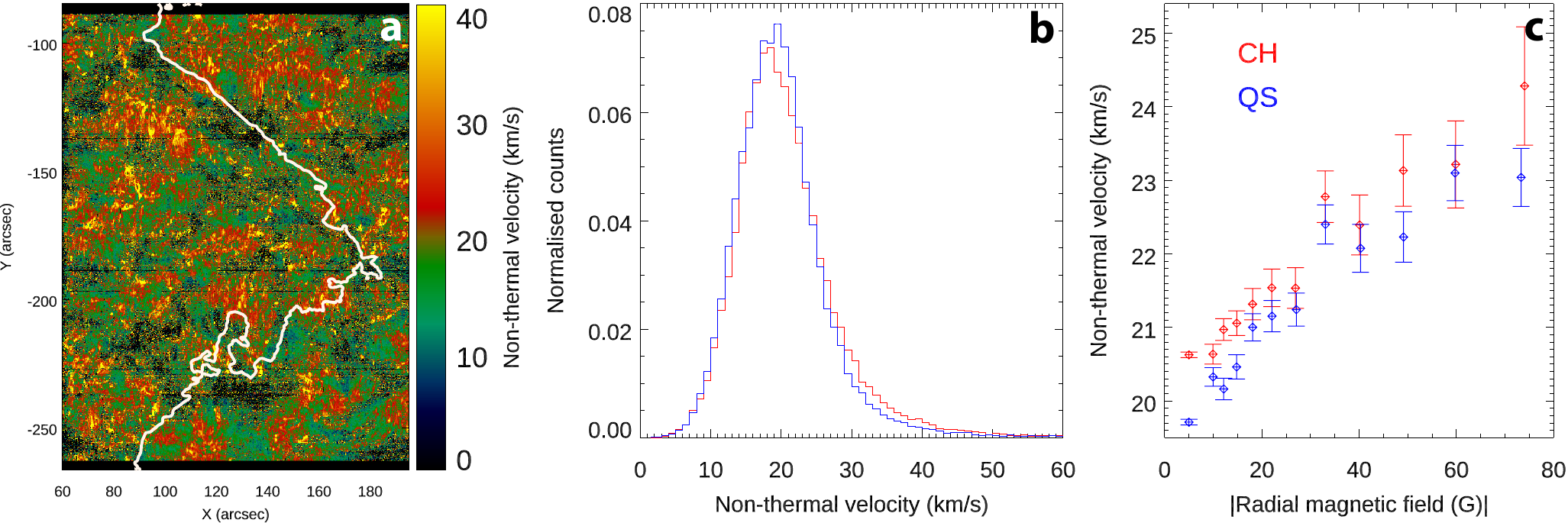}
\caption{Same as Fig.~\ref{fig:fig4} but for observations taken on 24-July-2014.}\label{fig:fig4_1}
\end{figure*}

\section{The dataset 3: observation on October 14, 2015}\label{sec:app_b}

\setcounter{figure}{0} \renewcommand{\thefigure}{B.\arabic{figure}} 
\begin{figure*}[h]
\centering
\includegraphics[width=0.95\textwidth]{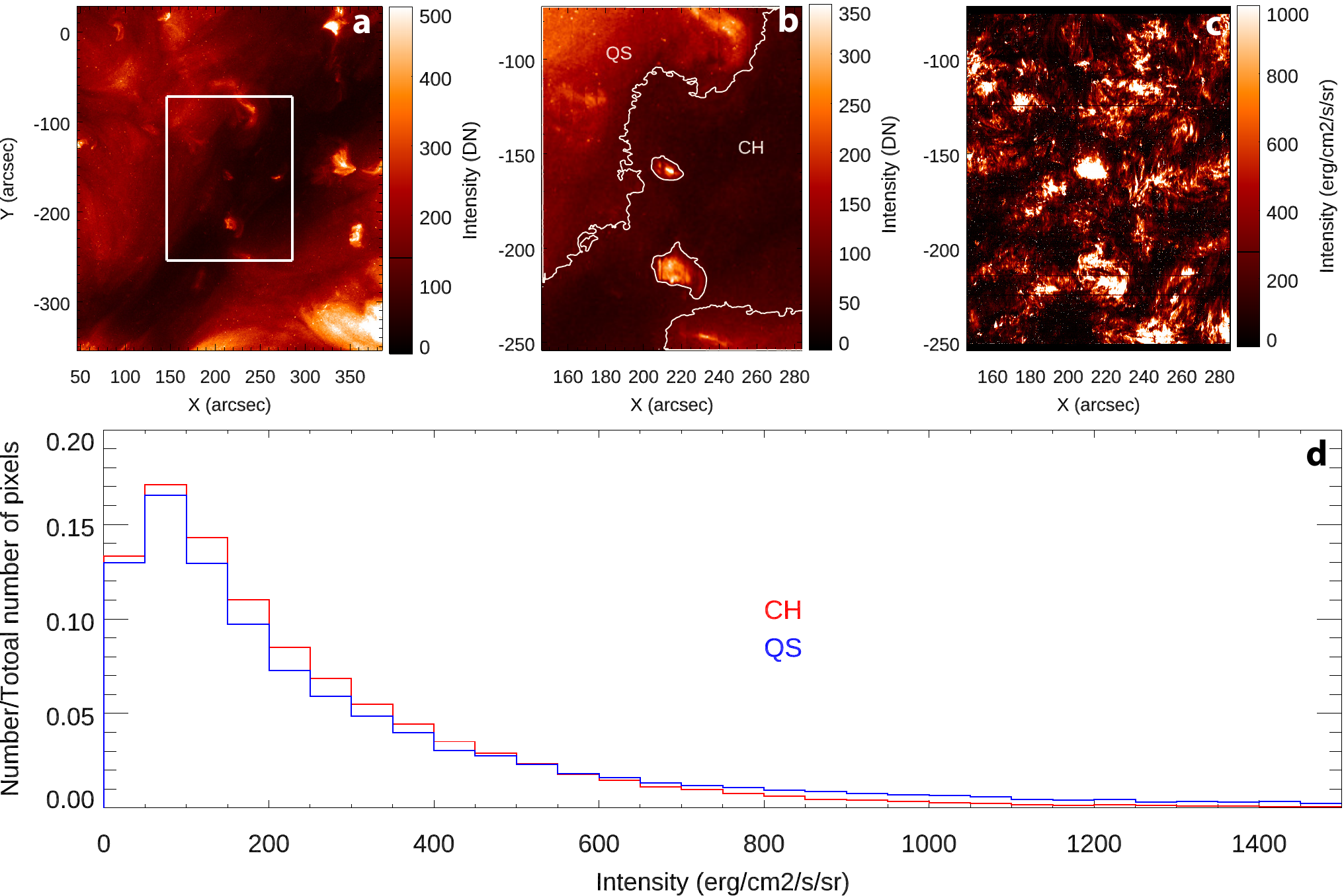}
\caption{Same as Fig.~\ref{fig:fig1} but for observations taken on 14-Oct-2015.}\label{fig:fig1_4}
\end{figure*} 
\begin{figure*}
\centering
\includegraphics[width=0.95\textwidth]{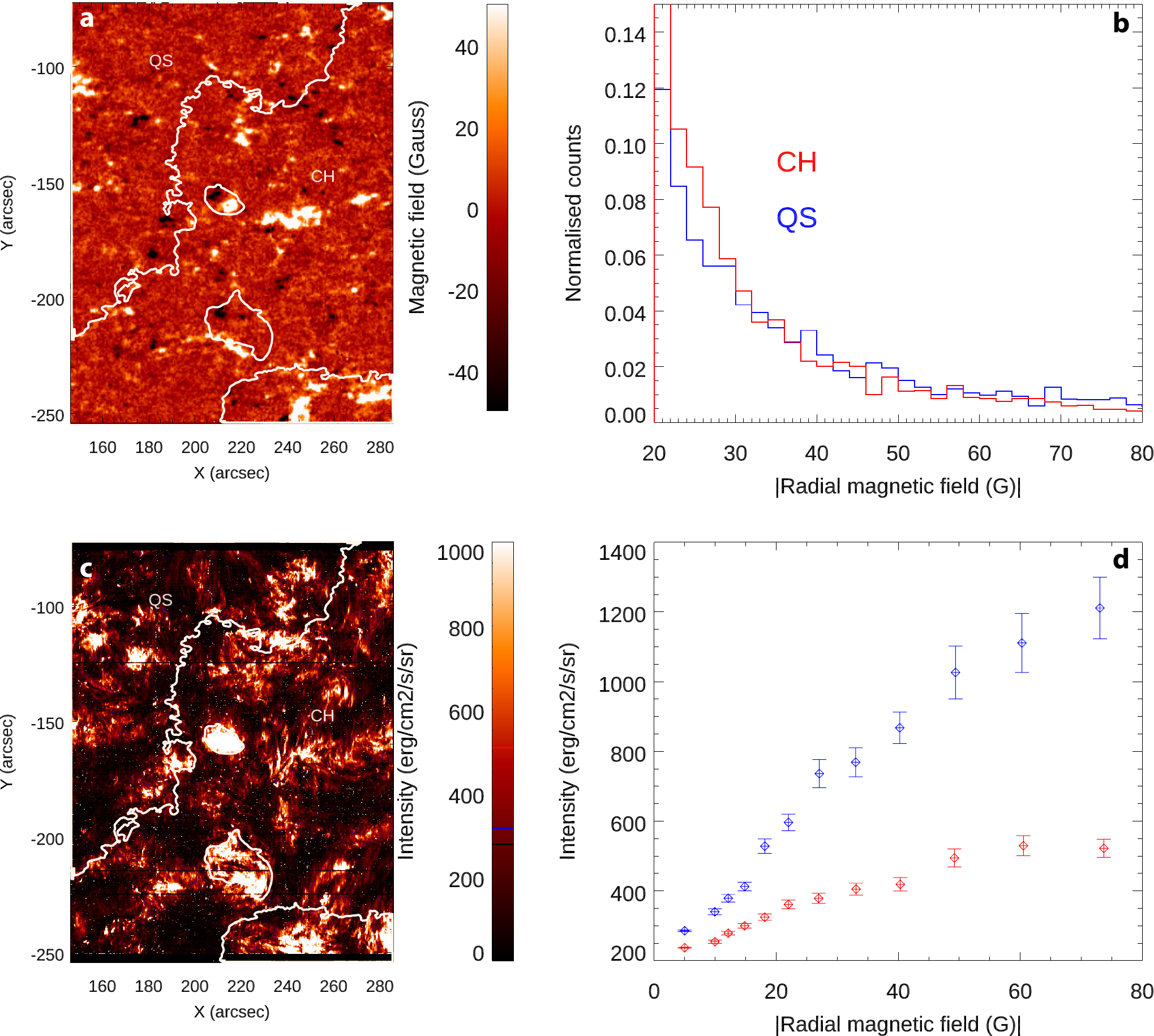}
\caption{Same as Fig.~\ref{fig:fig2} but for observations taken on 14-Oct-2015.} \label{fig:fig2_4}
\end{figure*}
\begin{figure*}
\centering
\includegraphics[width=0.95\textwidth]{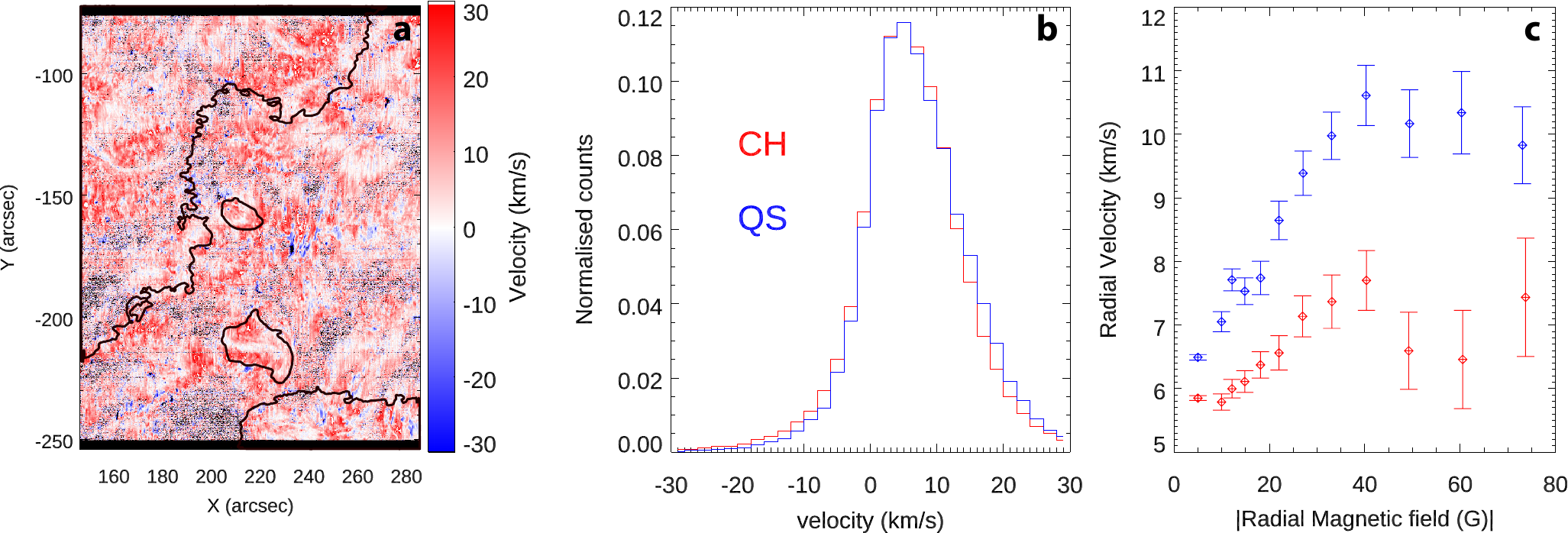}
\caption{Same as Fig.~\ref{fig:fig3} but for observations taken on 14-Oct-2015.}\label{fig:fig3_4}
\end{figure*}
\begin{figure*}
\centering
\includegraphics[width=0.95\textwidth]{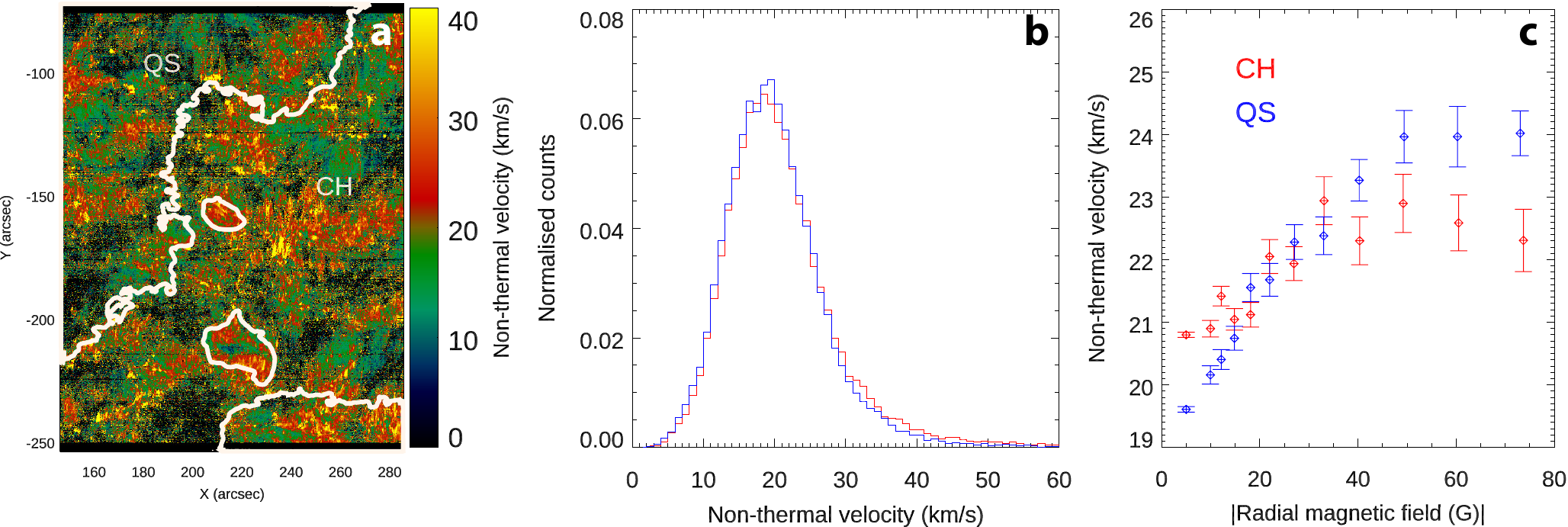}
\caption{Same as Fig.~\ref{fig:fig4} but for observations taken on 14-Oct-2015.}\label{fig:fig4_4}
\end{figure*}

\end{document}